\documentclass[twocolumn,prd,groupedaddress,nofootinbib,showpacs]
{revtex4}

\usepackage{latexsym}
\usepackage{amssymb}
\usepackage{graphicx}
\usepackage{dcolumn}
\usepackage{bm}

\begin{document}

\title{Lorentz-breaking effects in scalar-tensor theories of gravity}

\author{V. B. Bezerra$^{1a}$
C.N. Ferreira$^{2,3b}$ and
J. A. Helay\"el-Neto$^{3, 4c}$}

\affiliation{$^{1}$ Departamento de F\'{\i}sica, Universidade Federal da Para\'{\i}ba,
58059-970, Jo\~ao Pessoa, PB, Brazil
\\
$^{2}$ Instituto de F\'{\i}sica, Universidade Federal do Rio de Janeiro,
Caixa Postal 68528, 21945-910, Rio de Janeiro, RJ, Brazil
\\
$^{3}$ Grupo de F\'{\i}sica Te\'orica Jos\'e Leite Lopes, Petr\'opolis, RJ, Brazil
\\
$^4$ Centro Brasileiro de Pesquisas F\'{\i}sicas, Rua Dr. Xavier Sigaud 150;
Urca 22290-180, Rio de Janeiro, RJ, Brazil}

\date{\today}

\begin{abstract}
In this work, we study the effects of breaking Lorentz symmetry in scalar-tensor theories 
of gravity taking torsion into account. We show that a space-time with torsion 
interacting with a Maxwell field by means of a Chern-Simons-like term is able to 
explain the optical activity in syncrotron radiation emitted by cosmological distant radio 
sources. Without specifying the source of the dilaton-gravity, we study the 
dilaton-solution. We analyse the physical implications of this result in the Jordan-Fierz 
frame. We also analyse the effects of the Lorentz breaking in the cosmic string formation 
process. We obtain the solution corresponding to a cosmic string in the presence of 
torsion by keeping track of the effects of the Chern-Simons coupling and calculate the 
charge induced on this cosmic string in this framework. We also show that the resulting 
charged cosmic string  gives us important effects concerning the background radiation.The 
optical activity in this case is also worked out and discussed.

\end{abstract}


\maketitle

\section{Introduction}

The idea of the possible existence of extra dimensions of  space-time as proposed
in the Kaluza-Klein theory\cite{Klein} has inspired  the formulation of 
scalar-tensor theories of gravity. In these theories, the gravitational interaction
is mediated by one or several long-range fields, in addition to the usual tensor field
of the Theory of General Relativity and represents the simplest and most natural
generalization of General Relativity\cite{Fierz}-\cite{Damour}. 
The relevance of these scalar-tensor theories  resides in the fact that in the very early 
universe, or in the presence of strong gravitational fields, both the scalar and tensor 
aspects of gravity have to be taken into account. Nowdays,these effects are small but, 
to some extent, they can be observed. 

In addition to the scalar and tensor fields, we shall consider another one which may 
have an important role: the torsion field. This could influence some physical phenomena as, 
for example,  neutrino oscillations\cite{Adak}, and may have been an important element
in the early universe, when quantum effects of gravity were drastically important.

In this paper, we study the cosmic string configuration in the context of scalar-tensor 
theories of gravity\cite{CMH2001} with torsion\cite{VC2002,VCJA2003,Cris2002} and analyse 
their role in the geometric and topological features of the cosmic string solution. A 
cosmic string is what is called a topological defect and corresponds to a regular, 
classical solution to a gauge field theory which arises when a symmetry of the theory is 
spontaneously broken. In particular, in the framework of Cosmology, it may be generated 
during phase transitions in the early universe\cite{Vilenkin,Kibble}. The GUT defects 
carry a large energy density and, hence, are of interest in Cosmology, as potential 
sources  to explain the most energetic events in the Universe, like  the cosmological
gamma-ray bursts (GRBs)\cite{brandeb93}, ultra high energy cosmic rays (UHECRs) and very 
high energy neutrinos\cite{Brandenberger}, and gravitational-wave bursts\cite{B93,CHG2004} 

In recent years, there has ben a considerable interest in theories with the Lorentz and 
CPT violations. These theories may be implemented by a Chern-Simons type model in four 
dimensions.  In three dimensions, the Chern-Simons models have attracted a considerable 
attention due to the fact that the Maxwell-Chern-Simons-Higgs (MCSH) theory in a 
three-dimensional Minkowski space-time \cite{Paul} has some similarities with the theory 
of high-$T_C$ superconductivity. At large distances, the Chern-Simons term dominates over 
the Maxwell term and so it is reasonable to consider the simplest Abelian 
Chern-Simons-Higgs model, from which it was shown\cite{Hong} that there exists a vortex 
solution to the three-dimensional Abelian Chern-Simons-Higgs model, and an electrically 
charged vortex solution with a Chern-Simons term\cite{Schaposnik}. Motivated by these 
reasons, in our work we study the possibility to build up a cosmic string solution in the 
presence of a four-dimensional Chern-Simons type term \cite{Helayel2002}; in this 
dimension it presents a vector coupling, which can be indentified with the dilaton gradient 
in a scalar-tensor theory of gravity.

In a previous paper\cite{VHC2003}, only the effect of the interaction between the cosmic 
string-dilaton-solution in the background was considered; now, we analyse another aspect 
associated with the charge induced in the core of the string. An interesting application 
of a cosmic string with Chern-Simons-torsion coupling is to analyse the possible existence 
of a preferred direction in the sky. This subject has already discussed in the context of 
theories of gravity\cite{will93,HL00,HL01,das00,NR97,nr97-2} and observational 
cosmology\cite{birch82}. The idea is that the electromagnetic radiation travelling through 
the intergalactic medium interacts with its components and, if this radiation is initially 
plane polarized, the plane of polarization will rotate. Thus, if the Faraday rotation is 
taken into account, there is a residual rotation that does not vanish. Such a phenomenon, 
if it exists, would imply the violation of the Lorentz invariance\cite{will93}, with 
unpredictable consequences for fundamental physics\cite{grillo00}.

In the present paper, we consider that a plane polarized electromagnetic radiation has 
the plane of polarization rotated when it is travelling in the presence of the 
gravitational field generated by a screwed cosmic string in scalar-tensor theories of 
gravity. The motivation to consider such a background with torsion was already 
discussed\cite{kuhne97,capo99}. On the other hand, the assumption that gravity may be 
intermediated by a scalar field(or, more generally, by many scalar fields), 
in addition to the usual tensor field, has been considerably reassessed over the recent 
years. It has been argued that gravity may be described by a scalar-tensorial gravitational 
field, at least at sufficiently high energy scales.

This paper is organized as follows: In Section II, we present some aspects of 
scalar-tensor theories. In Section III, we introduce the Chern-Simons coupling to the free 
field theory. The cosmic string solution with Chern-Simons effects is presented in 
Section IV. In Section V, we discuss the possibility that this approach can be in 
accordance with all of the experiments where the optical activity appears. Finally, in 
Section VI, we present our Concluding Remarks.

\section{Scalar-tensor theories of gravity with torsion }

In this section, we set some results concerning the scalar-tensor theories of gravity 
with torsion. Let us consider that, in this case, gravity is represented by an action in 
the Jordan-Fierz frame\cite{Kim,Gaspperini}, which is given by

\begin{equation}
I=\frac 1{16\pi }\int d^4x\sqrt{{-\tilde g }}
\left[ {\tilde \phi}{ \tilde R }-\frac{\omega(\tilde \phi) }{\tilde  \phi}
\partial _\mu \tilde  \phi\partial^\mu \tilde \phi \right] +
I_m \label{acao1} \; ,
\end{equation}
where the matter action $I_m $ in Eq.(\ref{acao1}) is related with the dilaton-matter 
coupling. The function $ \omega $ in a general scalar-tensor theory has a $\tilde \phi $ 
dependence, but in the specific case of the Brans-Dicke theory it is a constant. The 
scalar curvature $\tilde R$, appearing in Eq.(\ref{acao1}) can be written as

\begin{equation}
\tilde R = \tilde R(\{\}) +
\delta \frac{\partial_{\mu} \tilde \phi \partial^{\mu} \tilde \phi}
{\tilde \phi^2} \; ,
\end{equation}
where $\tilde R(\{\})$ is the Riemann scalar curvature in the Jordan-Fierz frame and 
$\delta $ is the torsion coupling constant\cite{Gaspperini}. It is worthy to stress that 
in the scalar curvature $\tilde R $, the scalar function $\tilde \phi $ (the dilaton field) 
can act as a source of the torsion field. Therefore, in the absence of string spin, the 
torsion field may be generated by the gradient of this scalar field \cite{Kim}. In this 
case, the torsion can be propagated with the scalar field, and it can be written as

\begin{equation}
S_{\mu\nu}^{\hspace{.2 true cm} \lambda} =
( \delta^{\lambda}_{\mu} \partial_{\nu} \tilde \phi -
\delta^{\lambda}_{\nu} \partial_{\mu}\tilde \phi)/
2\tilde \phi. \label{torsion}
\end{equation}

The most general affine connection, $\Gamma_{\lambda \nu}^{\hspace{.3 true cm} \alpha}$, 
in this theory has a contribution arising from the contortion tensor, 
$K_{\lambda \nu}^{\hspace{.3 true cm}\alpha}$, as given below:

\begin{equation}
\Gamma_{\lambda \nu}^{\hspace{.3 true cm} \alpha} =
\{^{\alpha}_{\lambda \nu}\} +
 K_{\lambda \nu}^{\hspace{.3 true cm}\alpha} \; , \label{kont1}
\end{equation}
where the quantity  $ \{^{\alpha}_{\lambda \nu}\}$ is the Christoffel symbol computed 
from the metric tensor $g_{\mu \nu}$, and the contortion tensor, 
$K_{\lambda \nu}^{\hspace{.3 true cm}\alpha}$, can be written in  terms of the torsion 
field as

\begin{equation}
K_{\lambda \nu}^{\hspace{.3 true cm}\alpha} =
-\frac{1}{2}(S_{\lambda \hspace{.2 true cm} \nu}^{\hspace{.1 true
cm} \alpha} + S_{\nu \hspace{.2 true cm} \lambda}^{\hspace{.1 true
cm} \alpha} - S_{\lambda \nu}^{\hspace{.3 true cm} \alpha}) \; .
\end{equation}

Now, let us introduce a Maxwell-Chern-Simons coupling and analyse its consequences. Thus, 
we will consider the action for the matter in (\ref{acao1}), which we will indicate 
by $I_{MCS}$, as given by

\begin{widetext}
\begin{equation}
I_{MCS} =\int d^4x\sqrt{{-\tilde g }}
\left[ -\frac{1}{4} F_{\mu \nu}F^{\mu \nu} +
\frac{1}{3!} \lambda \varepsilon^{\mu \nu \alpha \beta}
{\tilde F}_{\mu \nu} A_\alpha {\tilde S}_\beta \right], \label{acao2} \; 
\end{equation}
\end{widetext}

\noindent
where we couple the electromagnetic dual field, $ {\tilde F}_{\mu \nu}$, and the vector 
potential, $A_\alpha$, to the torsion vector ${\tilde S}_\beta= 
- {\frac{\lambda}{3}} S_{\beta}$, which is responsible for the appearance of the preferred
cosmic direction, as suggested by the observations\cite{NR97}. The parameter $\lambda$ 
is the coupling constant of the theory, whose expected value will be estimated later on, 
using current astronomical data sets. In what follows, we shall investigate the role 
played by the Chern-Simons term in the scalar-tensor screwed cosmic string background.

In this paper, we shall consider the case where $S_{\mu}$ is a gradient of some scalar 
field,  $\phi$. This preserves gauge invariance but not the Lorentz invariance. In the 
scalar-tensor screwed cosmic string background, we may consider this scalar $\phi$ as the 
dilaton. Thus, using Eq.(\ref{torsion}) the torsion vector (in the Jordan-Fierz frame) is
defined as

\begin{equation}
\tilde{S}_{\mu} = \frac{3}{2} \partial_{\mu}\ln{\tilde \phi}\label{extra} \; .
\end{equation}

Although the action proposed in Eq.(\ref{acao1}) shows explicitly this scalar-tensor 
gravity feature, for technical reasons, we will adopt the Einstein (conformal) frame in 
which the kinematic terms of the scalar and tensor fields do not mix. In this frame, the 
action can be written as

\begin{equation}
\begin{array}{lrl}
{I} = \frac{1}{16\pi G} \int d^4x &\sqrt{-g}& \left[ R -
2g^{\mu\nu}\xi \partial_{\mu}\phi\partial_{\nu}\phi   \right] \\
& & \\
&+& I_{_{MCS}},
\end{array}
\label{acao3}
\end{equation}

\noindent 
where $I_{MCS}$ is the action of Maxwell-Chern-Simons given by (\ref{acao2}) with 
${\tilde S}_{\mu}$, interchanged by $S_{\mu}$ which is given by

\begin{equation}
 S_{\mu} = \frac{1}{2}\alpha(\phi)  \partial_\beta \phi ,
\end{equation}

\noindent
where $g_{\mu\nu}$ is written in the Einstein frame, $R(\{\})$ is the curvature scalar 
without torsion and $\xi$ is the parameter that includes the torsion contribution, 
$\delta $, so defined that

\begin{equation}
\xi = 1- 2\delta \alpha^2
\end{equation}

This more convenient formulation of the theory in terms of the gravitational field 
variables, $g_{\mu \nu}$ and $ \phi$, is obtained by means of the conformal transformation

\begin{equation}
\tilde{g}_{\mu\nu} = \Omega^2(\phi)g_{\mu\nu} \label{conform},
\end{equation}

\noindent
and by a redefinition of the quantity

\[G\Omega^2(\phi) = \tilde{\phi}^{-1}\; .\]

\noindent
This transformation puts into evidence that any gravitational phenomena will be affected 
by the variation of the gravitational {\it constant}, $G$, in the scalar-tensor gravity, 
a feature that is exhibited through the definition of a new parameter,

\[\alpha^2(\phi) \equiv \left( \frac{\partial \ln \Omega(\phi)}{\partial
\phi} \right)^2 = [2\omega(\tilde{\phi}) + 3]^{-1} ,\]

\noindent
which can be interpreted as the field-dependent coupling strength between matter and the 
scalar field. In order to make our calculations as general as possible, we will not fix 
the factors $\Omega(\phi)$ and $\alpha (\phi)$, leaving them as arbitrary functions of 
the scalar field.

In this context, the field equation of the eletromagnetic field becomes

\begin{equation}
\frac{1}{\sqrt{- g}} \partial_{\mu}\left[\sqrt{- g}
F^{\mu \nu}\right]= \lambda \alpha(\phi)\tilde
F^{\mu \nu}\partial_{\mu}\phi.\label{eletro0}
\end{equation}

For some purposes, it is more interesting to write down these equations of motion in 
terms of the electric and magnetic fields. Then, we consider the electric field, $E^i$, 
and magnetic field, $B^i$, defined as usual by:

\begin{equation}
E^i = F^{0i}, \hspace{2 true cm} B^i = - \epsilon^{ijk}F_{jk}.\label{not0}
\end{equation}

In what follows, let us consider a spatially flat isotropic Friedmann-Robertson-Walker(FRW) 
background,

\begin{equation}
ds^2 = a^2(\eta) (- d\eta ^2 + dx^2 +dy^2 +dz^2),
\label{frweq}
\end{equation}
where $\eta $ is the conformal time coordinate, defined by $d\eta = dt/a(t)$, $a(t)$ being the 
cosmological scale factor. Then, we have the following equations 

\begin{equation}
\begin{array}{rrl}
\vec{\nabla} \bullet \tilde{{\bf E}} &=& 
2\lambda \alpha(\phi_0)\vec{\nabla}\phi \bullet \tilde{{\bf B}},\\
\partial_{\eta}\tilde {\bf E} - \vec \nabla \times \tilde {\bf B} &= & 2\lambda \alpha(\phi)
[\partial_{\eta}\phi \tilde{{\bf B}}- \vec \nabla \phi \times \tilde {\bf E}],\\
\vec \nabla \times \tilde {\bf E} &=& - \partial_{\eta} \tilde{{\bf B}},\\
\vec \nabla \bullet \tilde{{\bf B}} &=& 0,
\end{array}\label{eb0}
\end{equation}

\noindent
where $\tilde{{\bf E}} = a^2 {\bf E}$ and $\tilde{{\bf B}} = a^2 {\bf B}$.

The equation of the motion for  dilaton field in this background is given by

\begin{equation}
(\partial_{\eta}^2 - \nabla^2) \phi + \frac{2}{a}\partial_{\eta} a \partial_{\eta} \phi = 
0,\label{phio}
\end{equation}
where $\phi $ is taken to be an arbitrary function of the space and time coordinates. Assuming that 
the solution for $\phi $ has the form

\begin{equation}
\phi(\eta, \vec{x}) = \phi_0(\eta) \cos({\vec{k}\bullet \vec{x}})
\end{equation}

\noindent
and substituting this into Eq.(\ref{phio}), we get the following equation

\begin{equation}
\partial_{\eta}^ 2 \phi + \frac{2}{a} \partial_{\eta} a \partial_{\eta} \phi + k^2 \phi =0.
\end{equation}

We can notice that, as consequence of this result, the overall homogenity of the universe over long 
distance scales is not disturbed by the inclusion of a spatial part in $\phi$.

Now, we study the modification introduced by this background in the Pointing vector. From the field
equations (\ref{eb0}), we derive the wave equations for the electric and magnetic fields, which take
the form that follows:

\begin{equation}
\begin{array}{lll}
(\partial_{\eta}^2 - \nabla^2) \tilde {\bf B} &= & 
-2 \nabla \times(\nabla \phi \times \tilde {\bf E}) \\
&&
2 \nabla \times (\dot \phi \tilde {\bf B}),
\end{array}
\end{equation}

\begin{equation}
\begin{array}{lll}
(\partial_{\eta}^2 - \nabla^2) \tilde {\bf E} &=& 2 \nabla \times (\dot \phi \tilde{ {\bf E}}-
2 \nabla \phi \times \dot{ \tilde {\bf E}})\\
&&  + \nabla (\nabla \bullet \tilde{ {\bf E}}) - 2 \ddot \phi \tilde{ {\bf B}}.
\end{array}
\end{equation}

It is easy to see that these equations reduce to the usual Maxwell equations 
whenever $\phi =0 $ or constant. From the equations given
in (\ref{eb0}), we find that

\begin{equation}
{\bf \nabla } \bullet {\bf S}  + \frac{\partial U}{\partial t} +\tilde{{\bf E}} \bullet \tilde{{\bf  J}}=0,
\label{S}
\end{equation}

\noindent 
where ${\bf S} = (\tilde {\bf E} \times \tilde {\bf B})$ is the Poynting vector and 
$U = \frac{1}{2}(\tilde {\bf E}^2 + \tilde {\bf B}^2)$ is the electromagnetic energy density. The 
presence of a term of the form $\tilde{{\bf E}} \bullet \tilde{{\bf  J}} $ indicates a sort of 
dissipation effect. It can be interpreted as the analogue of the Ohm's law, where the current is proportional to $\tilde {\bf B}$ and given by  $ \tilde {\bf J } = 2\lambda\alpha(\phi) \dot \phi \tilde {\bf B}$. This 
sort of current induced by $\tilde  {\bf B}$ is a feature of  models with Chern-Simons-type Lorentz 
breaking term.

\section{Screwed cosmic string model with Chern-Simons coupling}

\noindent

In this section, let us investigate the solution that corresponds to a cosmic string when the 
Chern-Simons coupling is included. We analyse the vortex regime of the fields. The action for 
screwed cosmic string, $I_{m}( \tilde g_{\mu\nu}, \Psi)$, in an Abelian Higgs model can be written as

\begin{equation}
I_m = I_{_{SCS}} + I_{_{MCSH}},
\end{equation}

\noindent
where  $I_{_{SCS}}$ is the action associated with a screwed cosmic string and $I_{MCSH}$ is the 
Maxwell-Chern-Simons-Higgs action. First, let us consider $I_{SCS}$, which can be written as 

\begin{equation}
\begin{array}{ll}
I_{_{SCS}} = \int d^4x \sqrt{{\tilde g}}\left[-\frac{1}{2}D_{\mu}\Phi
(D^{\mu}\Phi)^*- \frac{1}{4}H_{\mu
\nu}H^{\mu \nu} - V(|\Phi |) \right],
\end{array}
\label{acao4}
\end{equation}

\noindent
where $D_{\mu} \Phi=(\partial_{\mu} + iX_{\mu})\Phi$ is the covariant derivative. The field strength 
$H_{\mu \nu}$ is defined in the standard fashion, namely, $H_{\mu \nu}=\partial_{\mu}X_{\nu}-
\partial_{\nu}X_{\mu}$, with $X_{\mu}$ being the gauge field. The action given by Eq.(\ref{acao4}) 
has a $U(1) $  symmetry associated with the $\Phi$-field  and it is broken by the vacuum, giving rise 
to vortices of the Nielsen-Olesen type\cite{Nielsen}:

\begin{equation}
\begin{array}{ll} \Phi = \varphi(r )e^{i\theta} \; ,\\
X_{\theta} = \frac{1}{q}[P(r) - 1]& \\
X_t = X_t(r)& \mbox{time-like}.
\end{array}\label{vortex1}
\end{equation}

The boundary conditions for the fields $\varphi(r) $ and $P(r)$ are the
same as those for ordinary cosmic strings\cite{Nielsen}, namely,

\begin{equation}
\begin{array}{ll}
\begin{array}{ll}
\varphi(r) = \eta, & r \rightarrow
\infty, \\
\varphi(r) =0, & r = 0, \end{array}& \begin{array}{ll}
P(r) =0, & r \rightarrow \infty, \\
P(r) =1, & r= 0.
\end{array}
\end{array}
\end{equation}

The configuration for the other component, compatible with the cosmic string stability, is given by the 
following boundary conditions:

\begin{equation}
\begin{array}{ll}
\begin{array}{ll}
X_t(r) =0, & r \rightarrow \infty, \\
X_t(r) = b, & r= 0.  \end{array}
\end{array}
\label{config2}
\end{equation}

It is worthy to draw the attention to the fact that these components are important to study the
behaviour of the charges and currents.

The potential $V(\varphi, \sigma)$ triggering the spontaneous
symmetry breaking can be built in the most general case as

\begin{equation} V(\varphi) = \frac{\lambda_{\varphi}}{4} (
\varphi ^2 - \eta^2)^2 ,
\end{equation}

\vspace{.5 true cm}

\noindent
where $\lambda_{\varphi}$ is a coupling constant. This potential possesses also all the ingredients 
which yields  the formation of a cosmic string, in analogy with the ordinary cosmic string case, 
where $X_z = X_t = 0$ and without an external field.

The term $I_{_{MCSD}}$ is given by

\begin{equation}
\begin{array}{lrl}
I_{_{MCSH}}= \int d^4x &\sqrt{{\tilde g}}&\left[- \frac{\lambda_1 }{4}F_{\mu \nu}F^{\mu \nu} +
\frac{\lambda_2 }{2}
\varepsilon^{\mu \nu \alpha \beta}
H_{\mu \nu} X_\alpha S_\beta \right. \\
& & \\
&+ & \left.\frac{\lambda_3 }{2}
\varepsilon^{\mu \nu \alpha \beta}
H_{\mu \nu} Y_\alpha S_\beta  \right].
\end{array}\label{action5}
\end{equation}

In this action, we introduced the external field $F_{\mu \nu} $ that will be analysed in connection
with the Chern-Simons coupling. The parameters $\lambda_1 $, $\lambda_2 $ and $\lambda_3 $ will be 
analysed with respect to the charge effects.

Let us consider a cosmic string in a cylindrical coordinate system, $(t,r, \theta ,z)$, ($r \geq 0$ 
and $0 \leq \theta < 2 \pi $) defined in the Einstein frame. In this frame, we can write the metric 
for a time-like cosmic string as\cite{Patrick1,Mac}

\begin{equation}
ds^2 = e^{2(\gamma - \psi)}(dr^2 + dz^2 ) + \beta^2
e^{-2\psi}d\theta^2 - e^{2\psi}dt^2
\label{metric2},
\end{equation}

\noindent
where $\gamma, \psi$ and  $\beta$ depend only on $r$. We can find the relations between the parameters 
of the metric through the Einstein equations. Then, in the space-time with the metric defined in 
Eq.(\ref{metric2}) and considering that the components $t$ and $z$ of the dilaton vanishes outside 
the string, these equations become 

\begin{equation}
\beta'' = 8 \pi \tilde G \beta ( T^t_t + T^r_r) e^{2(\gamma -
\psi)}\label{eqq0}
\end{equation}

\begin{equation}
(\beta \gamma')' = 8 \pi \tilde G \beta ( T^r_r +
T^{\theta}_{\theta}) e^{2(\gamma - \psi)},\label{eqq1}
\end{equation}

\begin{equation}
(\beta \psi')'= 4 \pi \tilde G \beta ( T^t_t +
T^r_r + T^{\theta}_{\theta} -T^z_z)e^{2(\gamma -
\psi)},\label{eqq2}
\end{equation}
where $\prime$ denotes ''derivative with respect to $r$''.

The equation for the scalar field, in this background, is given by

\begin{equation}
\Box \phi = - 4 \pi \tilde G \alpha T. \label{eqq3}
\end{equation}

Notice, we are working with is more general than the one for the ordinary cosmic string.
This is so because we hope that the cosmic string has a superconductor character. This fact 
already happens in $(2+1)$-dimensions where it acquires charge by virtue of the Chern-Simons 
coupling\cite{Schaposnik}.

\subsection{The external structure }

Now, we study the behavior of the external solution in 
the case where the dilaton is 
time-dependent\cite{Emilia3} and a Chern-Simons 
coupling is taken into account. 
The solutions for $\beta(r)$ and $\gamma(r)$ are the 
same as obtained recently\cite{CMH2001} and read as 

\begin{equation}
\begin{array}{ll}
\beta = B r\\
\gamma = m^2 ln\left(\frac{r}{r_0}\right)
\end{array}
\end{equation}

In order to obtain the external solutions of the Eq.(\ref{eqq3}), we follow the same procedure 
already used\cite{VCJA2003}, but now the $\phi$-equation depends on radial coordinate $r$ 
as well as on the coordinates $z$ and $t$ (we assume rotational 
symmetry so that $\partial_{\theta} \phi =0$. In the external case, to garantee the 
stability at infinity of the solution, we have to impose that $\phi(t) \equiv \phi(z) 
\rightarrow 0$ for large distances from de defect, which means in practice for $r\geq 0$.
Then, the $\phi $ solution outside the string is the usual one

\begin{equation}
\phi'= \kappa^{-1} \frac{\lambda }{r}.
\end{equation}

\noindent
This implies that

\begin{equation}
R=2(\psi''+\frac{1}{\beta}\psi' - \phi'^2 +
\frac{m}{\beta^2})e^{2(\psi-\gamma)}=  2\phi'^2e^{2(\psi-\gamma)},
\end{equation}

\noindent
which is different from the result obtained in the framework of scalar-tensor theories of gravity 
without torsion\cite{VCJA2003}.

The external metric for the time-like screwed cosmic string takes, thus, the form

\begin{widetext}
\begin{equation}
ds^2 =  \left( \frac{r}{r_0} \right)^{-2n} W^2(r) \left[
\left( \frac{r}{r_0}\right)^{2m^2} (dr^2 +dz^2) +
B^2r^2d\theta^2 \right] - \left( \frac{r}{r_0} \right)^{2n}
\frac{1}{W^2(r)} dt^2 \label{m9},
\end{equation}
\end{widetext}

\noindent
with $W(r) = [(r/r_0)^{2n} + p]/[1+p]$ and the parameters 
$n, \lambda $ and $m$ are given by $n^2 = \kappa^{-1}\lambda^2 + m^2 $, 
with $\kappa^{-1}$ constant in the Brans-Dicke theory.

\subsection{Dilaton solution in the weak-field approximation}

Now, let us investigate the dilaton-torsion solution considering the 
weak field approximation\cite{Patrick1,Emilia3}.
To do this, we assume that the metric $g_{\mu \nu}$ 
and scalar field $\phi $ can be written as

\begin{equation}
\begin{array}{ll}
g_{\mu \nu} = \eta_{\mu \nu} + h_{\mu \nu},\\
\Omega(\phi) = \Omega_0 + \Omega'_0 \phi_{(1)},\\
T_{\mu \nu} = T_{(0) \mu \nu} + T_{(1) \mu \nu},\\
\phi = \phi_0 + \phi_{(1)},
\end{array}
\end{equation}

\noindent
where $\Omega(\phi_0) = \Omega_0 $ and the energy-momentum tensor in the time-like case, 
$\lambda_1 =1$, is given as follows:
\vspace{.5 true cm}

\begin{widetext}
\begin{eqnarray}
T^{t}_{(0) t} & = &  - \frac{1}{2}\Omega^{2}_0 \{
\varphi'^{2} +
\frac{1}{r^2}\varphi^2P^2 +  \Omega^{-2}_0(\frac{P'^2}
{r^2q^2})  + \varphi^2 X_t^2 + \Omega^{-2}_0 X_t'^2  + \Omega^{-2}_0 Y_t'^2  + 2\Omega^2_0 V \}  \\
&&\nonumber\\
&& \nonumber \\
T^{z}_{(0) z} & = & - \frac{1}{2}\Omega^2_0 \{
\varphi'^2  + \frac{1}{r^2}\varphi^2P^2 + \Omega^{-2}_0
(\frac{P'^2}{r^2 q^2})  - \varphi^2 X_t^2 - \Omega^{-2}_0 X_t'^2  - \Omega^{-2}_0Y_t'^2 +  2\Omega^2_0
V  \}\\
&&\nonumber \\
&&\nonumber \\
T^{r}_{(0) r} & = &  \frac{1}{2}\Omega^2_0 \{
\varphi'^2  - \frac{1}{r^2} \varphi^2P^2  + \Omega^{-2}_0
(\frac{P'^2}{r^2 q^2})+ \varphi^2 X_t^2 - \Omega^{-2}_0X_t'^2  -  \Omega^{-2}_0 Y_t'^2  - 2\Omega^2_0
V \} \\
&&\nonumber \\
&&\nonumber \\
T^{\theta}_{(0) \theta} & =  & - \frac{1}{2}\Omega^2_0 \{ \varphi'^2  -  
\frac{1}{r^2} \varphi^2P^2+ \Omega^{-2}_0
(\frac{P'^2}{r^2 q^2})  + \varphi^2 X_t^2+  \Omega^{-2}_0X_t'^2+
 \Omega^{-2}_0Y_t'^2 + 2\Omega^2_0
V\},
\end{eqnarray}
\end{widetext}

\noindent
and $\phi_0 $ is the constant dilaton value in the absence of the string. The equation for the 
dilaton can be written as

\begin{equation}
\Box \phi = \varepsilon {\cal T}(r),\label{phi3}
\end{equation}

\noindent
with $\varepsilon = 4 \pi G^* \alpha_0$ and ${\cal T} = - T$.

Up to the first order in $\varepsilon $, the equation for the dilaton field reads

\begin{equation}
\frac{\partial^2 {\phi}_1}{\partial t^2} - \frac{\partial^2 {\phi}_1}{\partial {z^2}} -
\frac{1}{r}\frac{\partial}{\partial r}\left(r \frac{\partial \phi_1}{\partial r} \right)
= {\cal T}(r),\label{equadilaton2}
\end{equation}

\noindent
with ${\cal  T}$ given by

\begin{widetext}

\begin{equation}
{\cal  T}(r) = \Omega^2_0\xi^{-1}\left( \varphi^{'2} + \frac{1}{r^2} \varphi^2 P^2 + 
\Omega_0^{-2}X_t'^2 +             \Omega^{-2}_0 Y_t^{'2} + 4 \Omega^2_0  V(\phi)\right)\label{func1}
\end{equation}

\end{widetext}

\noindent
where, at this point, we consider that the components which interact with the string are $t$ and  
$z$ ones. The scalar field, $\phi_{(1)}$, depends on the radial coordinate as well as on the 
coordinates $t$ and $z$, and is given by

\begin{equation}
\phi_{(1)}(t, r, z) = \chi(r) + f(r) \psi(z,t),\label{sol1}
\end{equation}

\noindent
where the function $f(r) $ is required to vanish outside the string core. This ansatz for the dilaton 
is not incompatible with the cosmic string solution.  Using solution (\ref{sol1}) in 
Eq.(\ref{equadilaton2}), we obtain

\begin{equation}
\left( \frac{\partial^2 \psi}{\partial t^2} - \frac{\partial^2 \psi}{\partial {z^2}} \right) 
\frac{1}{r}\frac{\partial}{\partial r}\left(r \frac{\partial \chi}{\partial r} \right)  -
\frac{1}{r}\frac{\partial}{\partial r}\left(r \frac{\partial f}{\partial r} \right)\psi =
{\cal T}(r).\label{equadilaton3}
\end{equation}

In the ansatz (\ref{sol1}), $\chi$ corresponds to a field which depends only on the radial coordinate. 
As ${\cal T}$ is the source of the pure radial component of the dilaton, let us assume that

\begin{equation}
\chi '' + \frac{1}{r}\chi ' =  {\cal T}(r) \label{chi}.
\end{equation}

Therefore, Eq.(\ref{equadilaton3}) turns into

\begin{equation}
\frac{1}{\psi}\left(\frac{\partial^2 \psi}{\partial t^2} - 
\frac{\partial^2 \psi}{\partial {z^2}} \right) = 
\frac{1}{f}\left( \frac{\partial^2 f}{\partial r^2} +
\frac{1}{r}{\frac{\partial f}{\partial r}}\right) = \hat \omega ,
\label{eqpsif}
\end{equation}

\noindent
with $\hat \omega = k^2 - \omega^2 $ being a constant. 
Thus, we have the following set of equations:
\begin{equation}
\frac{\partial^2 f}{\partial r^2} +
\frac{1}{r}{\frac{\partial f}{\partial r}} = \hat \omega f
\label{eqf}
\end{equation}

and 

\noindent
\begin{equation}
\frac{\partial^2 \psi}{\partial t^2} - \frac{\partial^2 \psi}{\partial {z^2}} = \hat \omega \psi.
\label{eqpsi}
\end{equation}

In order to have compatibility with the external solution, $f(r)$ must have the following behaviour
$\lim_{r \rightarrow \infty}f(r) =0$. Now, let us consider the space-like dilaton $\hat \omega = k^2$. 
In this case, we have that the solutions of Eqs.(\ref{eqf}) and (\ref{eqpsi}) are given, respectively, 
by the following expressions

\begin{equation}
f(r) = f_{I}I_0(kr) +f_k K_0(kr),
\label{sf}
\end{equation}

and

\begin{equation}
\psi = \psi_0\sin(kz). 
\label{sphi}
\end{equation}

Notice that the solutions (\ref{sf}) and (\ref{sphi}) depend only on the coordinates 
$r$ and $z$. For these the solutions, 
we have the same  interpretation already presented  
in the literature (\cite{Emilia3}). 

\section{The Charged cosmic string configuration}

Now, let us to study the cosmic string configuration considering the weak field approximation 
solution to the dilaton. In this case, we have analysed the Chern-Simons effects in the core of 
the string with $\lambda_1 = \lambda_3 =0 $ in (\ref{action5}). Thus, the equation 
of motion for the gauge field $X_{\mu}(r)$ is given by

\begin{equation}
\partial_{\alpha} H^{\alpha \beta} +
\lambda_2 \epsilon^{\nu \beta k \lambda}H_{\nu k}S_{\lambda} =j^{\beta}.
\end{equation}

If we use the screwed cosmic string ansatz, we find the following equations

\begin{equation}
\vec{\nabla } \times \vec{B}  - \lambda_2 \vec{S} \times \vec{E} = \vec{J}
\end{equation}

\noindent
and 

\begin{equation}
\vec{\nabla } \bullet \vec{E}- \lambda_2 \alpha(\phi_0) f(r)  \vec{S} \bullet \vec{B} = \rho.
\end{equation}

These equations are compatible with the asymptotic conditions for the string, i.e., 
outside it $S_{\lambda}$ vanishes. Based on 
discussions concerning the dilaton solution, the time-like gradient of 
the dilaton (\ref{extra}) is given, in linearized 
approximation, by

\begin{equation}
S_{\mu } = - \alpha(\phi_0) \partial_{z}\phi_{(1)}\delta_{\mu}^z =
-\alpha(\phi_0)f(r)S_z(kz)\delta_{\mu}^z,\label{S1}
\end{equation}

\noindent
where $S_z(kz) = k\psi_0 \cos(kz)$.
If we use the Gauss law, we find

\begin{equation}
\begin{array}{lll}
{\cal Q} &= & \int d^3 x j^0 = \frac{\lambda_2 }{2} \int d^3 x 
\epsilon^{ijk}H_{ij}S_k \\
&=& \pi \lambda_2 \alpha(\phi_0)S_z \int_0^{r_0} f(r) B(r) r dr .
\end{array}
\end{equation}

Other interesting equation is related with the internal electric field, 
in analogy with the London equation, and is given by

\begin{equation}
J = \int d^3 x j^{\theta} = \pi \lambda_2 \alpha(\phi_0) S_z\int_0^{r_0}  f(r) E r dr, \label{J5}
\end{equation}
where $E = H_{r t} $ is the eletric field inside the string.
Therefore, from the previous results, we can get the interesting 
conclusion that the screwed cosmic string in presence of the
Chern-Simons-Higgs coupling is charged.

\subsection{The external electric field  in a cosmic string backgrund with Maxwell-Chern-Simons-Higgs 
coupling}

In what follows, we consider the effects of the Chern-Simons term on the cosmic string gravitational 
field. To study the effects presented in the last section, let us consider the vector $S_{\mu}$ as 
space-like  as in (\ref{S1}). The configuration of the external field that interacts with the cosmic 
string is  $Y_{t}(r)$. In this situation, we have the following equation for the gauge field 
$X_{\mu}(r)$, 

\begin{equation}
\partial_{\alpha} H^{\alpha \beta} +
\epsilon^{\nu \beta k \lambda}(\lambda_2 H_{\nu k} + \lambda_3 F_{\nu k})S_{\lambda} =j^{\beta}.
\end{equation}

This equation is compatible with the asymptotic conditions of the string, i.e., outside the 
string $S_{\lambda}$ vanishes, because 
the $f(r)$ dependence. The equation that is equivalent to Eq.(\ref{J5}) 
and includes the external field is given by

\begin{equation}
J = \int d^3z j^{\theta} = \alpha(\phi_0)\pi S_z \int f(r)\hat E r dr,
\end{equation}

\noindent
where $\hat E = \lambda_2 E + \lambda_3 E_{ext}$, with $E_{ext}$ being the external electric field.

In the electric case, where we use the space-like torsion vector, we have that the external 
electromagnetic field that interacts with the string is  $Y_t$. The equation of motion for 
$X_{\mu}$ gives us

\begin{equation}
{\cal Q} =  \alpha(\phi_0)\lambda_2 \pi S_z \int_0^{r_0} f(r)B(r) r dr,
\end{equation}

\noindent
where we have considered $Q = 2\sqrt{2} \pi^2  \int_0^{r_0}f(r)B(r) r dr $.

Now, we put the external field $Y_{\mu}$ to interact with this charge. The equation of motion for 
this, with $\lambda_1=1$, is given by

\begin{equation}
\partial_{\mu} F^{\mu \nu} +
\lambda_3 \epsilon^{\nu \rho \alpha \beta} S_{\rho} H_{\alpha \beta} =0.
\end{equation}

\noindent

Proceeding as previously, we find that

\begin{equation}
E_{ext} = \epsilon(z) \frac{Q}{ \sqrt{2} \pi r},
\label{B2}
\end{equation}
where $\epsilon(z) = \lambda_2 \lambda_3 \alpha S_z $.
In the next section, let us study the energy-momentum content in the magnetic case, 
using the procedure of the electric case.

The energy per unit length, $U$, and the tension per unit length, $\tau $, are given,
respectively, by

\begin{equation}
\mu  = -2 \pi \int_0^{r_0} T^t_{_{(0)}t} r dr,
\end{equation}

\noindent

and

\begin{equation}
\tau = - 2\pi \int_0^{r_0} T^z_{_{(0)}z} r dr,
\end{equation}
where $ T^{\mu}_{_{(0)} \mu}$ represents the trace of the energy-momentum tensor without torsion.

The remaining components read as

\begin{equation}
\begin{array}{ll}
T_x = -2 \pi \int_{0}^{r_0} T^r_{_{(0)}r} r dr;\\
T_y = -2 \pi \int_{0}^{r_0}
T^{\theta}_{_{(0)}\theta} r dr.
\end{array}
\label{X}
\end{equation}

The energy conservation in the weak-field approximation reduces to

\begin{equation}
r\frac{dT^r_{(0) r}}{dr} = ( T^{\theta}_{(0) \theta} - T^r_{(0) r}).
\end{equation}

In order to find the metric, we use the Einstein-Cartan equations in the form $G^{\mu\nu}(\{\})~=
~8\pi G_0T^{\mu \nu}_{_{(0)}}$, in which the tensor $T_{_{(0)}\mu \nu}$(being first order in $G$) 
does not contain the torsion contribution due to the fact that we are working in the weak-field 
approximation and $\tilde G_0=\tilde \phi^{-1}~\equiv~G~\Omega^2_0$. After integration, we have

\begin{equation}
\int_0^{r_0}r dr \left(T_{(0) \theta}^{\theta } +
T^r_{(0) r} \right) = r_0^2 T^r_{(0) r}(r_0) = \frac{1}{2} r^2_0 Y_t^{'2}(r_0).
\end{equation}

Now, let us analyse the Lorentz breaking effects in the energy-momentum tensor. To do this, we use 
Eq.(\ref{B2}), which gives us

\begin{equation}
Y_t'  = \frac{\epsilon Q}{\sqrt{2} \pi r}
\end{equation}

Following the usual proceure, let us assume that the string is infinitely 
thin so that its energy-momentum tensor is given by

\begin{equation}  T^{\mu \nu}_{string} = diag [\mu , - \frac{(\epsilon Q)^2}{2 \pi} ,
- \frac{(\epsilon Q)^2}{2 \pi},-T]\delta(x)\delta(y),
\label{stensor2} \end{equation}

\noindent
where $ Q$ is the charge density in the core of the string. In this case, the energy-momentum tensor 
of the string source $T_{(0)\mu \nu}$ (in Cartesian coordinates)  has the following components

\begin{equation}
\begin{array}{ll}
T_{(0) tt} = \mu \delta(x)\delta(y) + \frac{(\epsilon Q)^2}{4 \pi }\nabla^2
\left(ln\frac{r}{r_0}\right)^2 ,\\
T_{(0) zz} = -\tau \delta(x)\delta(y) + \pi \frac{(\epsilon Q)^2}{4 \pi }\nabla^2
\left(ln\frac{r}{r_0}\right)^2,\\
T_{(0) ij} = - (\epsilon Q)^2 \delta_{ij}\delta(x)\delta(y)
+  \frac{(\epsilon Q)^2}{2 \pi} \partial_i \partial_j ln(r/r_0).
\end{array}
\end{equation}

Actually, these quatities are not conserved in the Einstein frame, but we know \cite{Emilia3} that
the $z$ contribution of the dilaton to the metric vanishes, because that the relevant quantity is 
the vacuum value and then the only contribution comes from the induced current. The linearized 
Einstein-Cartan equation in this case is 

\begin{equation}
\nabla^2h_{\mu \nu} = -16 \pi G<|(T_{(0) \mu \nu} - \frac{1}{2}\eta_{\mu \nu}T_{(0)}|>. \label{h}
\end{equation}

Then, calculating the expectation value, we get

\begin{equation}
\begin{array}{ll}
<|T_{(0) tt}|> = \mu \delta(x)\delta(y) + \frac{({\bar \epsilon} Q)^2}{ 4 \pi  }\nabla^2
\left(ln\frac{r}{r_0}\right)^2 ,\\
<|T_{(0) zz}|> = -\tau \delta(x)\delta(y) +
\pi \frac{({\bar \epsilon} Q)^2}{4 \pi}\nabla^2\left(ln\frac{r}{r_0}\right)^2,\\
<|T_{(0) ij}|> = - ({\bar \epsilon }  Q)^2 \delta_{ij}\delta(x)\delta(y)
+ \frac{({\bar \epsilon} Q)^2}{2 \pi} \partial_i \partial_j ln(r/r_0),
\end{array}
\end{equation}

\noindent
with $<S_z>$ given by

\begin{equation}
<S_z^2> = \psi_0^2 \int_{-\pi}^{\pi} \cos^2u du = \pi \psi_0^2 = S^2
\end{equation}

\noindent
and ${\bar \epsilon}^2 = \lambda^2 \gamma^2 S^2 \alpha^2   $,
where $u = kz $. Now we have to determine the solution of Eq.(\ref{h}). I order to do this let us 
follow the same procedure presented in \cite{VCJA2003,Patrick1} concerning to the supercontucting 
cosmic string.

\subsection{The metric solution  the in weak field approximation}

For a superconducting cosmic string\cite{VCJA2003}, the dynamics defined by Eq.(\ref{eletro}) is 
complicated by itself, but if we assume that we are very far from the source then the gravitational 
coupling can be neglected, and so we are left with

\begin{equation}
\chi = 2G_0
\alpha \xi^{-1} (\mu + \tau +
({\bar \epsilon}Q)^2) \ln \frac{\rho}{r_0}  \label{phi1} \; \label{phi}.
\end{equation}

The metric for a superconducting cosmic string in this case is considered in the linearized case as

\begin{equation}
ds^2 = (1-  h_{tt})[-dt^2 + a(t)^2(dx^2 + dy^2 + dz^2)],,\label{metric}
\end{equation}

\noindent
with the component  $h_{tt}$ in the Jordan-Fierz frame given by

\begin{widetext}
\begin{equation}
h_{tt} = -4 G_0\left\{ ({\bar \epsilon } Q)^2 \ln(\frac{\rho}{r_0}) +  \mu  -\tau -  
({\bar \epsilon }Q)^2
+ \frac{\alpha^2 \xi^{-1}}{2}(\mu + \tau +
({\bar \epsilon }Q)^2)\right\}\ln(\frac{\rho}{ r_0})\label{htt}.
\end{equation}
\end{widetext}

\section{Analysis of the Chern-Simons-like coupling in a scalar-tensor screwed cosmic string background  }

In this section, we study the cosmic string background in the context of a scalar-tensor theory 
including torsion with Lorentz breaking induced by the charge inside the string. We examine, 
in particular, the effect on the polarization of the synchroton radiation coming from cosmological 
distant sources, and the relation between the electric and magnetic components of the radiation. 
The radio emission from distant galaxies and quasars present the polarization vectors which are not 
randomly oriented, as naturally expected. This interesting phenomenon suggests that the space-time 
staying between the source and observer may exhibit some optical activity. In this case, we consider 
that torsion in the action (\ref{acao2}) is generated by the cosmic string, an important aspect 
analysed here, associated to the fact that the cosmic string is charged.

Replacing the linearized solution given by Eq.(\ref{phi}) into Eq.(\ref{acao2}), as we did in our 
previous paper \cite{VC2002}, we have

\begin{equation}
S_{\mu} = - 3\alpha \partial_{\mu}\chi\label{extra2}.
\end{equation}

In this context, the equation for the eletromagnetic field becomes

\begin{equation}
\frac{1}{\sqrt{-\tilde g}} \partial_{\mu}\left[\sqrt{-\tilde g}
F^{\mu \nu}\right]= \lambda \alpha(\phi_0)\tilde
F^{\mu \nu}\partial_{\mu}\chi.\label{eletro}
\end{equation}

For some purposes, it is  more interesting to write  down these
equations of the motion in terms of the electric and magnetic
fields. Then, we consider the electric field, $E^i$, and magnetic
field, $B^i$, defined as usual:

\begin{equation}
E^i = F^{0i} \hspace{2 true cm} B^i = - \epsilon^{ijk}F_{jk}\label{not1}
\end{equation}

Thus, Eq.(\ref{eletro}) can be written using the linearized solution given by (\ref{metric}).

Now, let us consider a FRW background given by (\ref{frweq}). Therefore, the equations of motion 
can be written as

\begin{equation}
\begin{array}{rrl}
\vec{\nabla} \bullet \tilde{{\bf E}}_{ext} &=
& 2\lambda \alpha \vec{\nabla}\chi \bullet \tilde{{\bf B}}_{ext}\\
\partial_{\eta}\tilde {\bf E}_{ext} - \vec \nabla \times \tilde {\bf B}_{ext} &= & -2\lambda \alpha
\vec \nabla \chi \times \tilde {\bf E}_{ext}\\
\vec \nabla \times \tilde {\bf E}_{ext} &=& - \partial_{\eta} \tilde{{\bf B}}_{ext}\\
\vec \nabla \bullet \tilde{{\bf B}}_{ext} &=& 0
\end{array}\label{eb1}
\end{equation}

Using the usual procedure\cite{VHC2003}, we find that the dispersion relation  
in powers of $S_{\alpha}$, to first order,
gives us
\begin{equation}
k_\pm = \omega \pm 2\lambda \xi^{-1} G_0 (\mu + \tau +
({\bar \epsilon }Q)^2) \hat{s} \cos({\gamma})\label{k} \; .
\end{equation}

\noindent

In this case, the parameter $Q$ is the charge in the vortex induced by 
Lorentz breaking and $\tau $ is the tension of the string.

The angle, $\beta$, between the polarization vector and the galaxy's major axis is defined as

\begin{equation}
<\beta> = \frac{1}{2}\frac{r}{\Lambda_s} \cos(\vec{k},\vec{s})\label{nr97},
\end{equation}

\noindent
where $< \beta >$ represents the mean rotation angle after Faraday's rotation is removed, $r$ is the 
distance to the galaxy, $\vec{k}$ the wave vector of the radiation, and $\vec{s}$ a unit vector.

The rotation of the polarization plane is a consequence of the difference in the propagation speed of 
the two modes, $\kappa_+$,$\kappa_-$, the main dynamical quantities computed above. This difference, 
defined as the angular gradient with respect to the radial (coordinate) distance, is expressed as

\begin{equation}
\frac{1}{2}(\kappa_+ - \kappa_-) = \frac{d\beta}{dr}\; ,
\end{equation}
where $\beta$ measures the specific entire rotation of the polarization plane, per unit length $r$, 
and is given once again by $\beta = \frac{1}{2}  \Lambda_s^{-1}  r \cos \gamma \;. $ In the case of 
the screwed cosmic string, the constant $\Lambda_s$, that encompasses the cosmic distance scale for 
the optical activity to be observed, can be written as a function of the cosmic string energy 
density $\mu $ as

\begin{equation}
\Lambda_s^{-1}= 2\lambda G_0
\alpha ^2 \xi^{-1} (\mu + \tau + ({\bar \epsilon }Q)^2)  \; .
\end{equation}

It is illustrative to consider a particular form for the arbitrary function $\alpha(\phi) =2 \times 10^{-2}$, corresponding to 
the Brans-Dicke theory. In this work we  use the Nodland and Ralston data to $\Lambda_s^{-1} = 
10^{-32} eV$ to obtain the estimative of the value of the torsion coupling constant to Chern-Simons 
theory $\lambda $. Using COBE data that 
$ 2G_0 \alpha (\phi_0)\xi^{-1} (\mu + \tau + ({\bar \epsilon }Q)^2)  \sim 10^{-6} $, 
we find that $\lambda \sim 10^{-26}$~eV.

Indeed, these results are in agreement with the recent measurement of optical polarization of light from
quasars and galaxies \cite{HL01,das00,NR97,jain99}. An interesting set of potential explanations of 
this effect has been put forward in Refs. \cite{Novikov,kuhne97,capo99,Sengupta}.

\section{Concluding Remarks}

In this work, we show that is possible to construct a cosmic string solution in the presence of a 
Chern-Simons coupling in the case where the Lorentz breaking vector is the dilaton gradient. Actually, 
supersymmetry imposes that for the Chern-Simons-type Lorentz breaking to be realized, the background 
vector that condensates must indeed be the gradient of a scalar field \cite{BBBCHN2004} this is why we 
adopted the torsion as the dilaton gradient.

There are very important consequences of this fact in the cosmic string solution, as for example, the 
existence of a charge induced  by the Lorentz breaking in the core of the string. Moreover, birrefringence effects may also show up. Both the external and internal solutions are consistent in the case of a 
time-like cosmic string. In the space-like  string, there are  problems concerning the 
solution of the dilaton equation, as already discussed\cite{Emilia3}. In order to solve this 
problem, it is necessary to introduce a massive scalar field. This case was not analysed in the 
present work, but it will be the subject of  a future investigation.
According to our present results, the background generated by this cosmic string is birrefringent and 
agrees with our previous analysis. The diferences between them is that, in the present 
case, the current has a scalar-tensor parameter which gives a damping in the produced effect; but, 
even in this situation, this effect could have been interesting consequences at the time which 
probably cosmic strings were formed.

\vspace{1 true cm}

{\large \bf Acknowledgments:}  The authors would like to thank (CNPq-Brazil) for
financial support. C. N. Ferreira also thank Centro Brasileiro de Pesquisas
F\'{\i}sicas (CBPF) for hospitality.

\end{document}